%% file: main.tex
\documentclass[letterpaper]{article} 
\usepackage[draft]{aaai2026}  
\usepackage{times}  
\usepackage{helvet}  
\usepackage{courier}  
\usepackage[hyphens]{url}  
\usepackage{graphicx} 
\usepackage{natbib}  
\usepackage{caption} 
\usepackage{algorithm}
\usepackage{algorithmic}

\usepackage{newfloat}
\usepackage{listings}
\DeclareCaptionStyle{ruled}{labelfont=normalfont,labelsep=colon,strut=off} 
\floatstyle{ruled}
\newfloat{listing}{tb}{lst}{}
\floatname{listing}{Listing}
\usepackage{amsmath,amssymb}
\usepackage{booktabs}
\usepackage{dirtree}
\usepackage{xcolor}
\usepackage[color=yellow,size=scriptsize]{todonotes}

\title{Generative AI Priors Increase Effective Sample Size in Clinical Trials}
\title{How many patients could we save with LLM priors?}
\author{
    Shota~Arai\textsuperscript{\rm 1},
    David~Selby\textsuperscript{\rm 2},
    Andrew~Vargo\textsuperscript{\rm 1}, 
    Sebastian~Vollmer\textsuperscript{\rm 1, 2}
}
\affiliations{
    \textsuperscript{\rm 1}Osaka Metropolitan University\\
    \textsuperscript{\rm 2}German Research Centre for Artificial Intelligence (DFKI)\\
    \textsuperscript{\rm 3}University of Kaiserslautern--Landau (RPTU)\\
    david.selby@dfki.de, shota.arai@dfki.de, sebastian.vollmer@dfki.de
}

\begin{document}

\maketitle

\begin{abstract}
Imagine a world where clinical trials need far fewer patients to achieve the same statistical power, thanks to the knowledge encoded in large language models (LLMs).
We present a novel framework for hierarchical Bayesian modeling of adverse events in multi-center clinical trials, leveraging LLM-informed prior distributions.
Unlike data augmentation approaches that generate synthetic data points, our methodology directly obtains parametric priors from the model.
Our approach systematically elicits informative priors for hyperparameters in hierarchical Bayesian models using a pre-trained LLM, enabling the incorporation of external clinical expertise directly into Bayesian safety modeling.
Through comprehensive temperature sensitivity analysis and rigorous cross-validation on real-world clinical trial data, we demonstrate that LLM-derived priors consistently improve predictive performance compared to traditional meta-analytical approaches.
This methodology paves the way for more efficient and expert-informed clinical trial design, enabling substantial reductions in the number of patients required to achieve robust safety assessment and with the potential to transform drug safety monitoring and regulatory decision making.
\end{abstract}

\begin{links}
    \link{Curated Datasets}{https://github.com/ybarmaz/bayesian-ae-reporting}
    \link{Raw Data (Project Data Sphere)}{https://data.projectdatasphere.org/projectdatasphere/html/content/104}
\end{links}

\section{Introduction}


Accurate modeling of adverse events (AEs) in clinical trials is crucial for drug safety assessment and regulatory decision-making \cite{hong2021meta}. Traditional approaches to AE modeling often struggle with limited sample sizes, heterogeneity across clinical sites, and the challenge of incorporating clinical expertise into statistical models. Hierarchical Bayesian models provide a principled framework for addressing these challenges by allowing information sharing across sites while accounting for site-specific variation.

The specification of prior distributions in Bayesian models remains a significant challenge, particularly in clinical contexts where domain expertise is abundant but difficult to quantify \cite{mikkola_prior_2024}.
Recent advances in large language models (LLMs) offer new opportunities for systematically eliciting expert knowledge and translating it into probabilistic priors \cite{selby_experts_2025}.
This paper explores the application of a pre-trained LLM for prior elicitation in hierarchical Bayesian modeling of individual patient data (IPD) from multi-center clinical trials. We systematically compare two representative LLMs: Llama 3.3 70B (Llama 3.3), a general-purpose language model, and MedGemma 27B (MedGemma), a model specifically fine-tuned for biomedical and clinical knowledge. Both models were used for prior elicitation and their impact on hierarchical Bayesian modeling was empirically evaluated.

We focus specifically on adverse event count modeling using a hierarchical Poisson--Gamma framework, where patient-level AE counts are modeled as Poisson distributed with site-specific rates following a Gamma distribution. The key innovation lies in using LLM-derived priors for the hyperparameters of this hierarchical structure, potentially improving model performance while incorporating clinical expertise.

Our empirical evaluation uses real clinical trial data from NCT00617669, a multi-center metastatic hormone-resistant prostate cancer (HRPC) study, providing a realistic testbed for methodology validation. Notably, this dataset was also used in \citet{barmaz_bayesian_2021}, allowing for direct comparison with established meta-analytical priors. Through systematic temperature sensitivity analysis and rigorous cross-validation, we demonstrate the practical utility of LLM-informed priors for clinical safety modeling, benchmarking their performance against the widely used meta-analytical prior framework. Crucially, our results reveal that LLM-based priors can enable substantial reductions in required sample size, demonstrating the potential of LLMs to reduce the number of patients required for statistically robust clinical trials.

\subsection{Related Work}

The application of LLMs as an interface to expert prior knowledge is an growing area of research.

Google has proposed agentic LLM frameworks to generate novel scientific hypotheses \cite{gottweis_ai-coscientist_2025}.
Such tools have been further augmented with external tools, including ability to access online literature databases and knowledge graph embeddings \cite{ghafarollahi_sciagents_2025}.
However, here we focus not on the qualitative outputs of these `AI co-scientists' but on the abilities of generative AI to assist with quantitative analysis tasks.

One prominent sub-field is data imputation and augmentation. Prior work has demonstrated the utility of LLMs for imputing tabular data \citep{nazir_chatgpt_2023}, spatio-temporal data \citep{wang_spatiotemporal_2025}, and in recommender systems \citep{ding_recommendation_2024}.
Researchers have also explored the impact of prompt design and data serialization techniques for this task \citep{srinivasan_prompt_2025, elkayal_reduction_2025}, as well as ensemble learning \citep{he_llm-forest_2025}.
\citet{isomura_llmovertab_2025} applied LLMs to oversampling imbalanced data.
In causal inference, \citet{huynh_improving_2025} used LLMs to generate synthetic counterfactual outcomes to improve conditional treatment effect estimation.
For a broader overview of LLM-based data augmentation, see \citet{ding_augmentation_2024}.

Another direction involves using LLMs for Bayesian prior elicitation.
\citet{selby_experts_2025} investigated the use of LLMs to increase effective sample size by eliciting parametric Bayesian conjugate priors for meteorological prediction.
This was extended by \citet{capstick_autoelicit_2025}, who built a mixture model from several Gaussian priors.
This contrasts with \citet{zhu_eliciting_2024}, who iteratively derive an implicit prior from synthetically generated data points.
Similarly, \citet{gouk_automated_2024} employed a data-augmentation approach to Bayesian prior elicitation, in which an LLM is used to generate synthetic data points that are used as pseudo-observations.
\citet{domke2025largelanguagebayes} proposed `Large Language Bayes', a framework that combines LLMs with probabilistic programming languages (e.g. Stan) to generate a candidate set of models, which are then averaged to produce a final posterior distribution.

\subsection{Contributions}

In the present paper, we build upon these advances to apply LLM-based prior elicitation to the high-stakes domain of clinical trials.
Unlike synthetic data augmentation apporaches that generate additional trainign examples, our methodological directly improves the statistical model through informative priors, potentially offering greater transparency and interpretability.
We provide a systematic methodology for LLM-based prior elicitation, employing distinct prompting strategies (blind and disease-informed) and conducting a comprehensive temperature sensitivity analysis to ensure robustness.

We specifically focus on the benefits of `expert' knowledge to reduce the number of participants required in clinical trials to obtain effective treatment effect estimates.

The contributions of this paper are as follows:
\begin{enumerate}
    \item We demonstrate that LLM-elicited priors consistently outperform traditional meta-analytical approaches in hierarchical Bayesian modeling of adverse events, achieving the best performance with Llama 3.3 using blind prompts at higher temperature settings ($T=1.0$).
    \item We provide empirical evidence that LLM-informed priors enable substantial sample size reduction in clinical trials while maintaining predictive accuracy---our experiments show that using only 80\% of training data with LLM priors achieves comparable or better performance than meta-analytical approaches with full data.
    \item We establish a systematic methodology for temperature sensitivity analysis and prompt strategy evaluation in LLM-based prior elicitation, revealing that general clinical expertise encoded in LLMs is sufficient for effective prior specification without requiring disease-specific prompting.
\end{enumerate}

\section{Hierarchical Bayesian AE Modeling with LLM-Informed Priors}

\subsection{Model Formulation}
We consider individual patient data (IPD) from multi-center clinical trials, where each patient $i$ in site $j$ has an observed adverse event count $y_{ij}$. Our hierarchical Bayesian model is structured as follows:
\begin{align}
y_{ij} &\sim \text{Poisson}(\lambda_j) \quad \text{(Patient-level likelihood)} \\
\lambda_j &\sim \text{Gamma}(\alpha, \beta) \quad \text{(Site-specific AE rates)} \\
\alpha &\sim \pi_\alpha(\cdot) \quad \text{(Hyperprior for shape)} \\
\beta &\sim \pi_\beta(\cdot) \quad \text{(Hyperprior for rate)},
\end{align}
where $\lambda_j$ represents the expected AE rate for site $j$, and $\alpha, \beta$ are hyperparameters governing the distribution of site-specific rates. The key innovation lies in the specification of hyperpriors $\pi_\alpha(\cdot)$ and $\pi_\beta(\cdot)$ using LLM-derived clinical expertise.

\subsection{LLM-Based Prior Elicitation}
We develop a systematic approach for eliciting hyperpriors using Llama 3.3, and MedGemma, leveraging its training on extensive clinical literature. Our methodology involves two distinct prompting strategies:

\textbf{Blind Prompting:} We query each LLMs without providing specific disease context (see Listing~\ref{lst:blind-prompt}), relying on its general clinical trial expertise:

\begin{listing}[tb]
\caption{Prompt for LLM-based Blind Prior Elicitation}
\label{lst:blind-prompt}
\begin{lstlisting}
You are a biostatistics expert specializing in clinical trials and Bayesian analysis.

TASK: Provide ONLY rate parameters for exponential priors in a hierarchical Bayesian model.

Model: 
- Each patient i in site j has AE count: y_ij ~ Poisson(lambda_j)
- Site-specific rates: lambda_j ~ Gamma(alpha, beta)  
- REQUIRED: alpha ~ Exponential(rate_alpha), beta ~ Exponential(rate_beta)

IMPORTANT:
- Use your expert knowledge and draw on published clinical trials, empirical data, or established domain knowledge to set informative (not weakly-informative or non-informative) prior rates.
- Avoid using vague or default values. Base your answer on realistic clinical data or strong prior experience relevant to typical AE rates in multi-center trials.

RESPOND WITH EXACTLY THIS JSON FORMAT (no markdown, no backticks, no other text):
{
"alpha_rate": number,
"beta_rate": number
}

Note: Exponential(rate) has mean = 1/rate. Rate must be > 0.
\end{lstlisting}
\end{listing}

\textbf{Disease-Informed Prompting:}  
The disease-informed prompt is identical to the blind prompt except for the following changes:
\begin{itemize}
    \item The introductory sentence is replaced with:\\
    ``You are a biostatistics expert specializing in oncology clinical trials and Bayesian analysis.''
    \item The following additional section is inserted before the model specification:
    \begin{lstlisting}
Clinical Context:
- Disease: Metastatic hormone-resistant prostate cancer (HRPC)
- Treatment: Control arm (placebo/standard care)
- Population: Adult oncology patients
- Study: Multi-center RCT
    \end{lstlisting}
    \item All instructions regarding data sources and prior information refer specifically to ``HRPC control arms in multi-center oncology trials'' rather than general clinical trial data.
\end{itemize}

The full disease-informed prompt is provided in the Supplement.

Both prompting strategies request exponential priors of the form $\alpha \sim \text{Exponential}(\lambda_\alpha)$ and $\beta \sim \text{Exponential}(\lambda_\beta)$, with the LLM providing specific rate parameters.

\subsection{Temperature Sensitivity Analysis}
To ensure robust prior elicitation, we systematically vary the GPT-4 temperature parameter across $T \in \{0.1, 0.5, 1.0\}$:
\begin{itemize}
\item ${T=0.1}$: Low temperature for consistent, focused responses
\item $T=0.5$: Moderate temperature balancing consistency and diversity  
\item ${T=1.0}$: Higher temperature for creative, varied responses
\end{itemize}

For each temperature-prompt combination, we conduct 5 independent LLM queries per cross-validation fold and aggregate the resulting parameters using arithmetic mean, ensuring statistical stability.

\subsection{Baseline Comparison}
We compare LLM-derived priors against established meta-analytical priors from \citet{barmaz_bayesian_2021}:
$$\alpha \sim \text{Exponential}(0.1), \quad \beta \sim \text{Exponential}(0.1)$$
These represent current best practices for hierarchical AE modeling in clinical trials, providing a rigorous benchmark for evaluation.

\section{Empirical Evaluation}

\subsection{Dataset}
Our evaluation uses real individual patient data from the control arm of the multi-center clinical trial NCT00617669 (metastatic hormone-resistant prostate cancer, HRPC), which is publicly available via the Project Data Sphere platform. The dataset was originally curated and described by \citet{barmaz_bayesian_2021}, and includes adverse event counts for 468 patients across 125 clinical sites, providing a realistic testbed for hierarchical AE modeling. Patient-level AE counts range from 0 to 140, with substantial heterogeneity across sites (mean site size: 3.74 patients, range: 1-27). Notably, this dataset is identical to that used in \citet{barmaz_bayesian_2021}, enabling direct methodological comparison to established work.

The hierarchical structure of the data makes it particularly suitable for evaluating site-level information borrowing and the impact of hyperprior specification on model performance. Access to the de-identified data via Project Data Sphere ensures that our methodology addresses practical challenges in clinical safety monitoring and supports reproducible research.

\subsection{Experimental Design}
We employed two main experimental designs to evaluate Bayesian prior elicitation for clinical trial data analysis using Large Language Models (LLMs):

\begin{enumerate}
    \item \textbf{Cross-Validation Model Comparison:} 
    We compared the predictive performance of LLM-generated Bayesian priors (Llama 3.3, MedGemma) against a meta-analytical baseline. For each method, we performed 5-fold stratified cross-validation at the site level, ensuring balanced representation of site sizes (small, medium, large) across folds. The hierarchical Poisson-Gamma model was fit using Markov Chain Monte Carlo (MCMC) with 1000 post-warmup draws and 1000 warmup iterations across 4 parallel chains. LLM priors were elicited under two prompting strategies (Blind and Disease-Informed) and three temperature settings (0.1, 0.5, 1.0), with five independent queries per condition averaged for robustness. Predictive performance was evaluated using Log Predictive Density (LPD).

    \item \textbf{Sample Efficiency Analysis:}
    To assess sample efficiency, we focused on the best-performing LLM and compared its performance to the meta-analytical baseline as the amount of training data was varied. The data was split into 70\% training and 30\% testing at the site level, and the training set was subsampled at 20\%, 40\%, 60\%, 80\%, and 100\% levels. For each subsample, Bayesian models were fit and evaluated on a consistent held-out test set, with 20 independent replications per condition. The same hierarchical Poisson-Gamma framework and evaluation metrics were used as in the cross-validation experiment.
\end{enumerate}

Both experiments employ a common evaluation framework based on out-of-sample predictive performance. 
Log Predictive Density (LPD) serves as the primary evaluation metric, calculated at the patient level using posterior predictive distributions.
The use of LPD as a primary metric for Bayesian predictive model comparison follows established practices in medical statistics, where LPD provides a robust summary of posterior predictive accuracy~\citep{boulet2019bayesian}.
For each test patient with observed AE count $y_{\text{obs}}$, we computed the LPD as follows:
\begin{equation}
    \mathrm{LPD} = \log \mathbb{E}_{\text{posterior}} \left[ P(y_{\text{obs}} \mid \lambda_{\text{new}}) \right],
    \label{eq:lpd-def}
\end{equation}
where $\lambda_{\text{new}} \sim \mathrm{Gamma}(\alpha, \beta)$ represents the expected AE rate for a new site, with $\alpha$ and $\beta$ drawn from their posterior distributions.

The expectation in Equation~\ref{eq:lpd-def} was approximated via Monte Carlo sampling. 
For each posterior sample $(\alpha^{(i)}, \beta^{(i)})$, we sampled $\lambda_{\text{new}}^{(i)} \sim \mathrm{Gamma}(\alpha^{(i)}, \beta^{(i)})$ and computed the Poisson log-probability $\log P(y_{\text{obs}} \mid \lambda_{\text{new}}^{(i)})$. 
The final LPD was then obtained using the log-sum-exp trick for numerical stability:
\begin{equation}
    \mathrm{LPD} = \log \left( \frac{1}{S} \sum_{i=1}^{S} \exp\left( \log P(y_{\text{obs}} \mid \lambda_{\text{new}}^{(i)}) \right) \right),
    \label{eq:lpd-mc}
\end{equation}
where $S$ is the number of posterior samples.

This design allows for a comprehensive comparison of LLM-based prior elicitation methods and their efficiency relative to traditional meta-analytical approaches.

\subsubsection{LLM Interaction for Prior Elicitation}
\begin{itemize}
    \item \textbf{LLM Choice:} Prior elicitation was performed using two large language models: Llama 3.3 (general-purpose) and MedGemma (biomedical-specialized). Meta-analytical priors were also included for baseline comparison.
    \item \textbf{Prompting Strategy:} For each parameter, a structured prompt was constructed according to the Blind or Disease-Informed protocol, with explicit instructions for output format and context. Prompts were identical for both LLMs except for disease context specification.
    \item \textbf{Input Variables to LLM:} Description of the clinical context (if applicable), the specific parameter to elicit, and the required output format (JSON).
    \item \textbf{Output Variables from LLM:} Numeric values for prior parameters (e.g., rate for Exponential prior for alpha and beta).
    \item \textbf{Implementation Note:} For each temperature and prompt setting, five independent queries were performed and outputs were averaged. All interactions and parsing were handled via Python scripts, ensuring robust error handling and reproducibility.
\end{itemize}

\section{Results}

\subsection{Experiment 1: Cross-Validation Model Comparison}

\subsubsection{Cross-Validation Methodology}
Our cross-validation approach follows established best practices for hierarchical data structures \citep{hastie2009elements}. We performed 5-fold stratified cross-validation at the site level to ensure balanced representation across different site characteristics. Sites were stratified into three groups based on patient count using tercile cutoffs (33.33\% and 66.67\%), creating small, medium, and large site categories. This stratification prevents potential bias from uneven site size distribution in train/test splits and ensures each fold contains representative samples from all site types.

This site-level splitting respects the hierarchical structure of multi-center clinical trials and provides a realistic evaluation scenario where models must predict AE rates for entirely new clinical sites.

\subsubsection{Performance Comparison}
Each model was evaluated under Blind and Disease-Informed prompting strategies and three temperature settings (0.1, 0.5, 1.0). For each temperature-prompt combination, we conducted 5 independent LLM queries per cross-validation fold and aggregated the resulting parameters using arithmetic mean to ensure statistical stability. Predictive performance was assessed using the common evaluation framework described above.

\subsubsection{Summary Table}
Table~\ref{tab:cv_summary} summarizes the LPD results for each model, prompt type, and temperature setting. The best performing model was Llama 3.3 with Blind prompt at $T=1.0$ (LPD = -3.842 $\pm$ 0.294). 
\begin{table*}[htbp]
    \centering
    \begin{tabular}{llrr}
    \toprule
    Model & Prompt Type & Temperature & LPD ($\pm$ SD) \\
    \midrule
        Llama 3.3 & Blind & 0.1 & -3.963 $\pm$ 0.464 \\
        Llama 3.3 & Blind & 0.5 & -3.959 $\pm$ 0.468 \\
        Llama 3.3 & Blind & 1.0 & -3.842 $\pm$ 0.294 \\
        Llama 3.3 & Disease-Informed & 0.1 & -4.063 $\pm$ 0.640 \\
        Llama 3.3 & Disease-Informed & 0.5 & -3.948 $\pm$ 0.450 \\
        Llama 3.3 & Disease-Informed & 1.0 & -3.909 $\pm$ 0.409 \\
        MedGemma & Blind & 0.1 & -4.023 $\pm$ 0.566 \\
        MedGemma & Blind & 0.5 & -4.133 $\pm$ 0.680 \\
        MedGemma & Blind & 1.0 & -3.844 $\pm$ 0.308 \\
        MedGemma & Disease-Informed & 0.1 & -4.121 $\pm$ 0.689 \\
        MedGemma & Disease-Informed & 0.5 & -3.997 $\pm$ 0.536 \\
        MedGemma & Disease-Informed & 1.0 & -3.971 $\pm$ 0.517 \\
        Meta-analytical & - & - & -3.963 $\pm$ 0.464 \\
    \bottomrule
    \end{tabular}
    \caption{Summary of LPD for each model, prompt type, and temperature (Experiment 1).}
    \label{tab:cv_summary}
\end{table*}

\subsubsection{Prior Parameter Analysis}
\begin{figure}[htbp]
    \centering
    \includegraphics[width=0.8\linewidth]{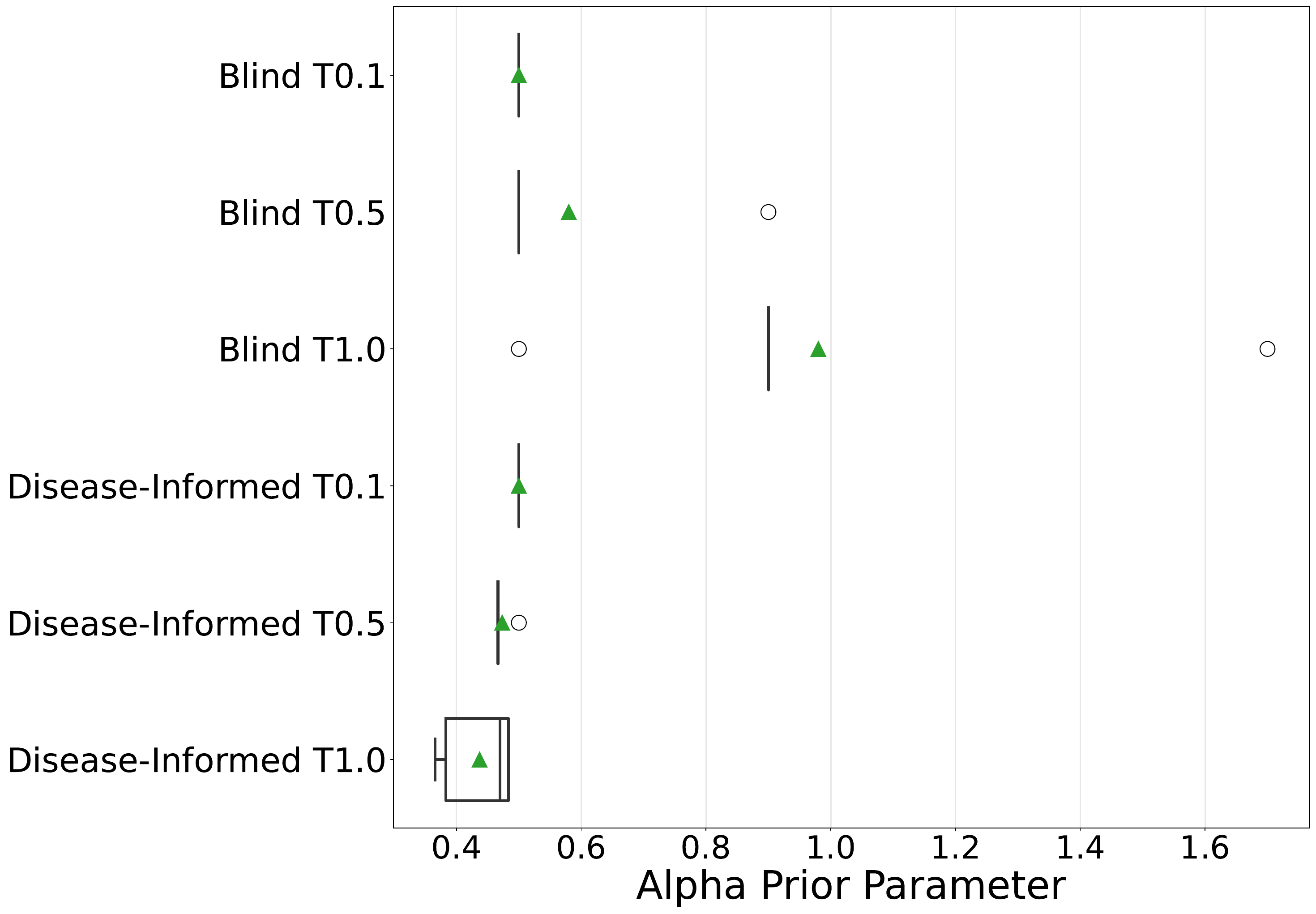}
    \caption{Boxplot of elicited alpha prior parameters for Llama 3.3 across prompt types and temperatures.}
    \label{fig:llama3.3_alpha_parameter_boxplot}
\end{figure}

\begin{figure}[htbp]
    \centering
    \includegraphics[width=0.8\linewidth]{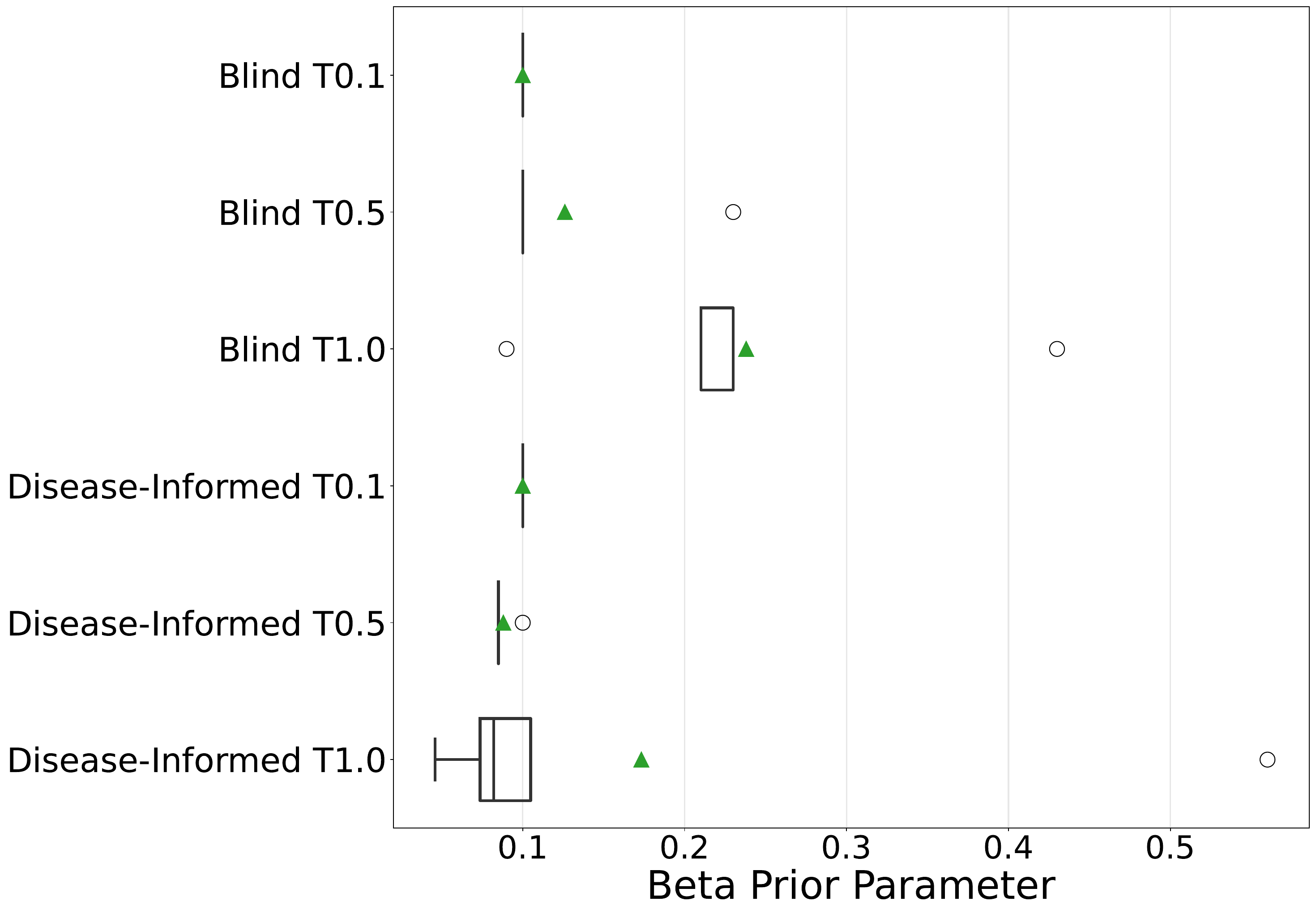}
    \caption{Boxplot of elicited beta prior parameters for Llama 3.3 across prompt types and temperatures.}
    \label{fig:llama3.3_beta_parameter_boxplot}
\end{figure}

\begin{figure}[htbp]
    \centering
    \includegraphics[width=0.8\linewidth]{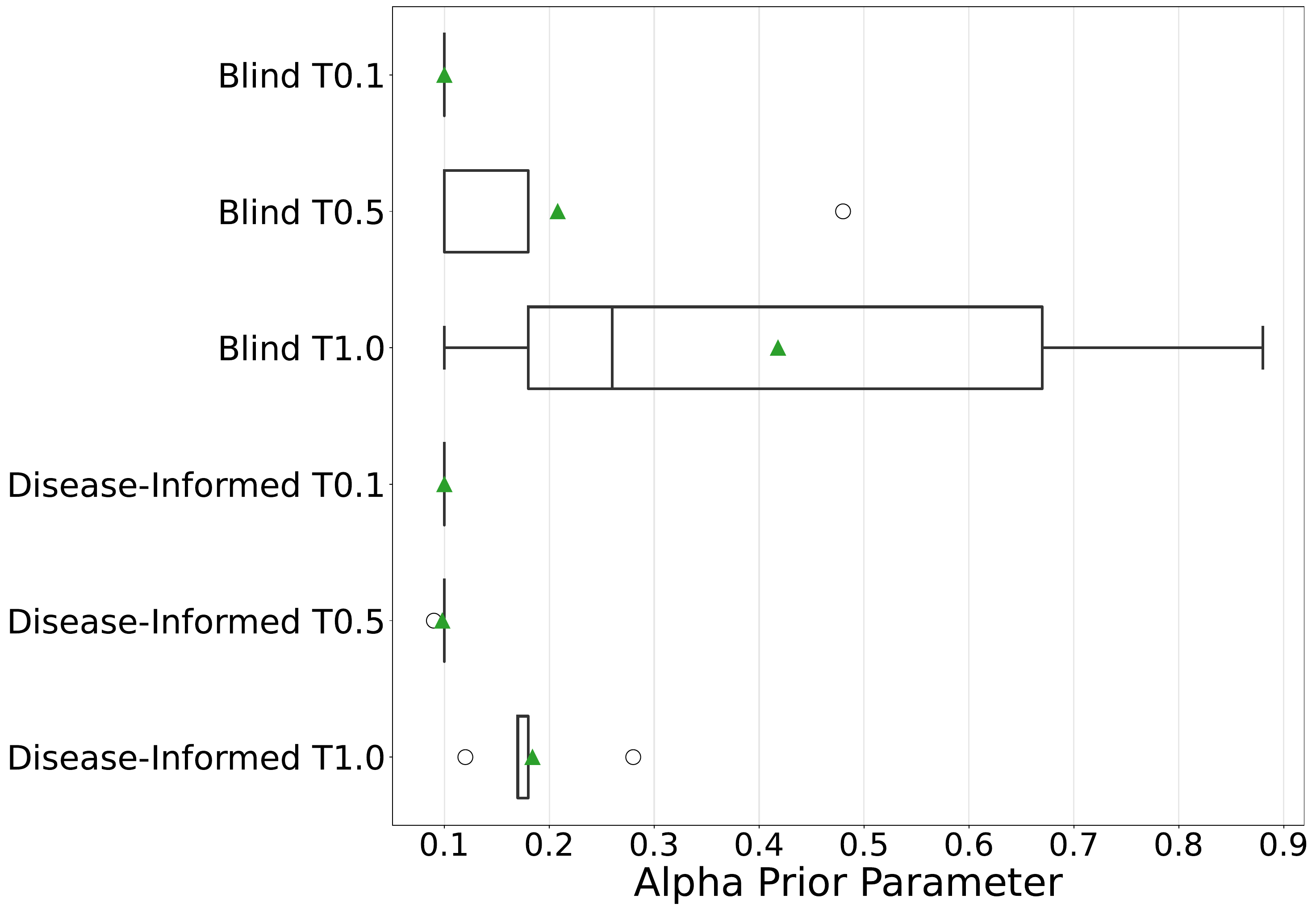}
    \caption{Boxplot of elicited alpha prior parameters for MedGemma across prompt types and temperatures.}
    \label{fig:medgemma_alpha_parameter_boxplot}
\end{figure}

\begin{figure}[htbp]
    \centering
    \includegraphics[width=0.8\linewidth]{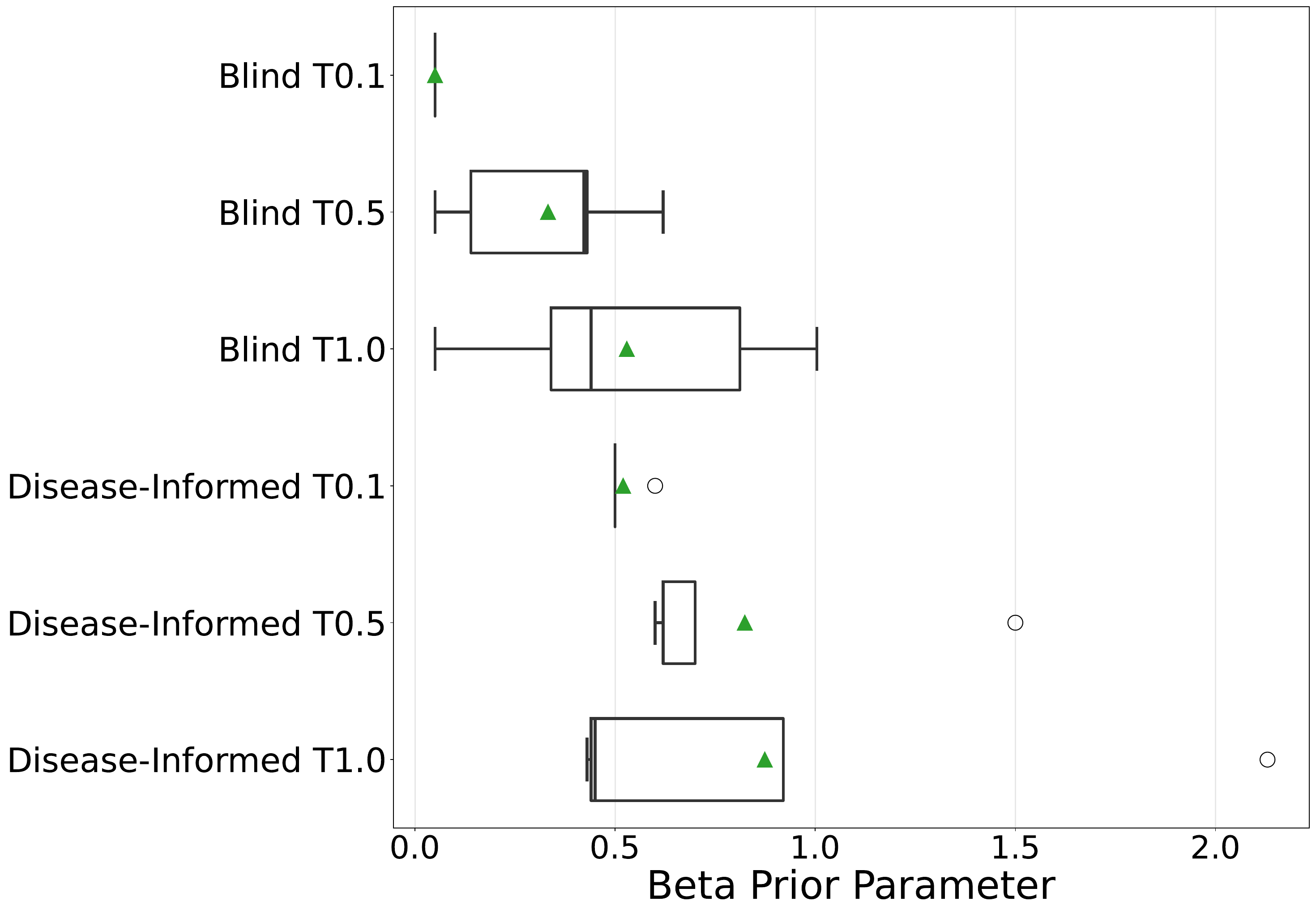}
    \caption{Boxplot of elicited beta prior parameters for MedGemma across prompt types and temperatures.}
    \label{fig:medgemma_beta_parameter_boxplot}
\end{figure}

Figures~\ref{fig:llama3.3_alpha_parameter_boxplot}, \ref{fig:llama3.3_beta_parameter_boxplot}, \ref{fig:medgemma_alpha_parameter_boxplot}, and \ref{fig:medgemma_beta_parameter_boxplot} show the distributions of elicited alpha and beta prior parameters for Llama 3.3 and MedGemma under different prompting strategies and temperature settings.

For Llama 3.3, the Blind $T=0.1$ and Disease-Informed $T=0.1$ conditions produced highly consistent priors (zero variance), while higher temperatures ($T=0.5$, $T=1.0$) resulted in increased variability, especially for the beta parameter. The Blind $T=1.0$ condition showed the largest spread (Alpha SD: 0.438, Beta SD: 0.122), indicating that higher temperature settings lead to more diverse prior suggestions. Disease-Informed $T=1.0$ also showed increased variance, particularly for beta (SD: 0.217).

MedGemma exhibited greater overall variability in both alpha and beta parameters compared to Llama 3.3. The Blind $T=0.1$ and Disease-Informed $T=0.1$ conditions were again highly consistent, but variance increased substantially at higher temperatures, especially for beta (Blind $T=1.0$ Beta SD: 0.380, Disease-Informed $T=1.0$ Beta SD: 0.732). Disease-Informed prompts tended to produce higher beta values and greater spread than Blind prompts.

Comparing the two models, Llama 3.3 produced higher alpha values (Mean: 0.578, SD: 0.258) and lower beta values (Mean: 0.138, SD: 0.109) than MedGemma (Alpha Mean: 0.185, SD: 0.182; Beta Mean: 0.522, SD: 0.451). This suggests that Llama 3.3 tends to elicit more concentrated priors for site-level rates, while MedGemma's priors are more dispersed, especially for beta.

Overall, temperature increases led to greater prior parameter variability for both models, with MedGemma showing more pronounced effects. These differences in prior parameter distributions may influence the degree of information borrowing and model regularization in downstream Bayesian analysis.

\subsubsection{Key Findings}
The best performing model was Llama 3.3 (Blind, $T=1.0$). All LLM-based priors outperformed the meta-analytical baseline.

\subsection{Experiment 2: Sample Efficiency Analysis}

\subsubsection{Training/Test Data Splitting and Subsampling}
Based on the results from Experiment 1, we focused on the best-performing configuration (Llama 3.3 with Blind prompts at $T=1.0$) to assess sample efficiency. Data splitting was performed at the site level rather than patient level to respect the hierarchical structure of multi-center clinical trials. We implemented a 70:30 train/test split using stratified sampling, where each stratum (small, medium, large sites based on patient count) contributed proportionally to both training and testing sets.

This approach ensures that the test set represents a realistic scenario where new clinical sites with unknown characteristics need AE rate prediction. Training data was used exclusively for model fitting and hyperparameter estimation, while test data served only for out-of-sample performance evaluation. To assess sample efficiency, the training set was systematically subsampled at 20\%, 40\%, 60\%, 80\%, and 100\% levels while maintaining stratified representation across site sizes.

\subsubsection{Performance Comparison}
For each training subsample size, we conducted 20 independent replications to ensure robust statistical inference. Each replication used a different random seed for site sampling while maintaining the same fixed test set for consistent evaluation. LPD was calculated on the held-out test set using the common evaluation framework, allowing direct comparison of sample efficiency between LLM-derived and meta-analytical priors.

\subsubsection{Summary Table}
Table~\ref{tab:sample_efficiency} summarizes the LPD results for each training sample size. Llama 3.3 maintained superior performance even with reduced training data, achieving comparable or better LPD than the meta-analytical baseline at full data.

\begin{table*}[htbp]
    \centering
    \begin{tabular}{lrr}
    \toprule
    Training Sample Size & LPD Mean & LPD Std \\
    \midrule
    $20\%$ & $-4.186$ & 0.196 \\
    $40\%$ & $-4.148$ & 0.157 \\
    $60\%$ & $-4.108$ & 0.125 \\
    $80\%$ & $-4.093$ & 0.107 \\
    $100\%$ & $-4.094$ & 0.124 \\
    Meta-analytical (100\%) & $-4.103$ & 0.136 \\
    \bottomrule
    \end{tabular}
    \caption{Sample efficiency analysis: LPD for Llama 3.3 (Blind, $T=1.0$) at different training sample sizes (Experiment 2).}
    \label{tab:sample_efficiency}
\end{table*}

\subsubsection{Key Findings}
Llama 3.3 (Blind, $T=1.0$) achieved similar or better predictive performance than the meta-analytical baseline, even when using substantially less training data. This demonstrates the sample efficiency advantage of LLM-derived priors in hierarchical Bayesian modeling for clinical trials.

\section{Discussion}
Our results demonstrate that LLM-elicited priors, specifically those generated by Llama 3.3 and MedGemma, can improve predictive performance in hierarchical Bayesian modeling of adverse events in clinical trials. The cross-validation experiments showed that Llama 3.3 (Blind, $T=1.0$) achieved the best Log Predictive Density (LPD), outperforming both MedGemma and the meta-analytical baseline. Notably, all LLM-based priors provided superior predictive accuracy compared to traditional meta-analytical approaches.

The prior parameter analysis revealed that Llama 3.3 tends to elicit more concentrated priors for site-level rates, while MedGemma's priors are more dispersed, especially for the beta parameter. Temperature sensitivity analysis indicated that higher temperatures ($T=1.0$) lead to greater variability in prior parameters, particularly for MedGemma, which may affect the degree of regularization and information borrowing in the Bayesian model. Disease-Informed prompts did not consistently outperform Blind prompts, suggesting that general clinical expertise encoded in LLMs is sufficient for effective prior elicitation in this context.

The sample efficiency experiment further demonstrated that Llama 3.3-derived priors maintain strong predictive performance even when the amount of training data is substantially reduced, achieving comparable or better LPD than the meta-analytical baseline at full data. Specifically, our analysis reveals that using only 80\% of the training data with LLM-informed priors (LPD = -4.093) outperforms the meta-analytical baseline using 100\% of the training data (LPD = -4.103). This 20\% reduction in required training data translates to approximately 66 fewer patients needed in our study context (from 328 to 262 patients in the training set), representing significant cost savings and reduced patient burden in clinical trial execution.

These findings suggest that LLM-based prior elicitation is a promising approach for improving statistical efficiency and predictive accuracy in clinical trial analysis. The ability to systematically incorporate clinical expertise through LLMs offers a practical solution for enhancing model performance, especially in settings with limited sample sizes or heterogeneous site distributions. The demonstrated sample efficiency gains have important implications for trial design, potentially enabling smaller trials to achieve the same statistical power as larger traditional studies, thereby accelerating drug development timelines and reducing costs.

Limitations of this study include the focus on a single disease area (HRPC) and dataset, as well as the use of only two LLM architectures. Future research should explore validation across diverse clinical contexts, comparison of additional LLMs, and formal assessment of prior informativeness and clinical impact. Integration with regulatory frameworks and extension to other types of clinical endpoints are also important directions for further work.

\section{Conclusion and Future Work}
This study demonstrates that LLM-informed prior elicitation, using models such as Llama 3.3 and MedGemma, can substantially improve predictive performance and sample efficiency in hierarchical Bayesian modeling for clinical trials. Through rigorous cross-validation and sample efficiency experiments on real HRPC trial data, we showed that LLM-derived priors consistently outperform traditional meta-analytical baselines, even when the available training data is limited. Llama 3.3 (Blind, $T=1.0$) provided the best overall results, and both LLMs enabled robust inference with fewer patients, highlighting the practical value of generative AI for clinical safety analysis.

Our prior parameter analysis revealed that LLMs can flexibly generate priors with varying degrees of concentration and regularization, depending on temperature and prompt strategy. While Blind and Disease-Informed prompts both yielded effective priors, higher temperature settings increased parameter variability, especially for MedGemma. These findings suggest that prompt and temperature tuning can be leveraged to control prior informativeness and model behavior in Bayesian analysis.

Importantly, we also considered the potential synergy between LLM-based prior elicitation and LLM-driven data augmentation, as discussed in the related work section. Combining these approaches may further enhance predictive accuracy and model robustness, especially in small-sample or heterogeneous clinical settings. Future research should systematically evaluate the joint use of LLM priors and synthetic data generation and develop formal criteria to evaluate their combined impact on Bayesian inference.

The limitations of this work include the focus on a single disease area and dataset and the use of only two LLM architectures. Further validation across diverse clinical domains, exploration of additional LLMs, and integration with regulatory frameworks are important next steps. Extending the methodology to other types of clinical endpoints and investigating the interpretability and acceptance of LLM-driven Bayesian models in real-world decision-making contexts will also be critical for broader adoption.

In summary, LLM-based prior elicitation represents a promising approach to improving statistical efficiency and predictive performance in clinical trial analysis. By enabling comparable levels of statistical power with fewer participants, this approach has the potential to reduce the number of patients required in future trials. Its combination with LLM-based data augmentation also represents a promising direction for future research, with the potential to transform the design and analysis of clinical studies.

\section{Relationship to Prior Work}
Our work builds upon recent advances in leveraging large language models for Bayesian prior elicitation and clinical decision support. While previous studies such as \citet{selby_experts_2025} explored LLM-based prior elicitation in healthcare contexts, our work is the first to systematically apply this approach to hierarchical adverse event modeling in multi-center clinical trials.

The closest related work includes \citet{barmaz_bayesian_2021}, who established meta-analytical priors for clinical trial contexts, providing the baseline comparisons used in our evaluation. Our approach extends this foundation by incorporating LLM-derived clinical expertise, demonstrating superior predictive performance while maintaining the interpretability advantages of explicit prior specification.

In contrast to synthetic data augmentation approaches that generate additional training examples, our methodology directly improves the statistical model through informative priors, offering greater transparency and potential regulatory acceptance. This distinction is particularly important in clinical contexts where model interpretability and validation are critical for regulatory approval and clinical adoption.


\section*{Appendices}

Further information on the experimental design, implementation details and computational requirements is given in the supplementary materials.

\bibliography{references}

\input{supplement}

\end{document}

%% file: supplement.tex
\appendix
\counterwithin{figure}{section}

\section*{\LARGE Supplementary materials}

\vspace{1em}

\section{Experiment Design and Replicates}

\paragraph{Number of Replicates} To ensure statistical robustness and account for model stochasticity, each experimental configuration was evaluated multiple times. For cross-validation experiments, we used 5-fold CV with 5 LLM queries per fold. For sample efficiency experiments, we performed 20 independent LLM queries per sample size to obtain robust estimates.

\paragraph{Sample Size Variation} For the sample efficiency experiment, multiple subsampling ratios ($\rho \in \{0.2, 0.4, 0.6, 0.8, 1.0\}$) were used to simulate different levels of data availability.

\paragraph{Controlled Comparison} For each cross-validation fold, three primary conditions were tested:
    \begin{enumerate}
        \item Hierarchical Bayesian model with \textbf{meta-analytical priors} (baseline).
        \item Hierarchical Bayesian model with \textbf{LLM-elicited informative priors} using blind prompts.
        \item Hierarchical Bayesian model with \textbf{LLM-elicited informative priors} using disease-informed prompts.
    \end{enumerate}


\section{Implementation Details}
\subsection{Model Fitting}
All Bayesian models were fitted using Markov Chain Monte Carlo (MCMC) with the following specifications:
\begin{itemize}
    \item \textbf{Sampling:} 1000 post-burn-in iterations with 1000 burn-in samples
    \item \textbf{Chains:} 4 parallel chains for convergence diagnostics
    \item \textbf{Convergence:} Gelman-Rubin $\hat{R} < 1.1$ for all parameters
    \item \textbf{Software:} PyMC implementation with automatic differentiation
\end{itemize}

\subsection{Cross-Validation Design}
\begin{itemize}
    \item \textbf{Stratification:} Sites stratified by patient count terciles (small: $\le$2, medium: 3-4, large: $\ge$5 patients)
    \item \textbf{Fold Composition:} 24 sites per fold (12 small + 6 medium + 6 large), averaging 88 patients per fold
    \item \textbf{Reproducibility:} Fixed random seeds for fold assignment and MCMC initialization
    \item \textbf{Evaluation:} Out-of-sample predictions on held-out sites using posterior predictive distributions
\end{itemize}

\section{LLM Interaction Protocol}
\subsection{Query Implementation}
\begin{itemize}
    \item \textbf{API:} Meta Llama 3.3 70B (llama-3.3-70b-instruct) and MedGemma 27B (medgemma-27b-it) via Chat AI Academic Cloud
    \item \textbf{Rate Limiting:} Automatic retry with exponential backoff for API limits
    \item \textbf{Response Parsing:} JSON format validation with error handling for malformed responses
    \item \textbf{Aggregation:} For cross-validation: 5 independent queries per fold and temperature setting. For sample efficiency: 20 independent queries per sample size
\end{itemize}

\subsection{Example LLM Responses}

Representative examples of LLM prior rate outputs (one for each setting):

\begin{itemize}
    \item \textbf{LLaMA 3.3 (llama-3.3-70b-instruct):}
        \begin{itemize}
            \item Blind prompt: \\
                  \texttt{\{"alpha\_rate": 0.5, "beta\_rate": 0.1\}}
            \item Disease-informed prompt: \\
                  \texttt{\{"alpha\_rate": 0.5, "beta\_rate": 0.1\}}
        \end{itemize}
    \item \textbf{MedGemma 27B (medgemma-27b-it):}
        \begin{itemize}
            \item Blind prompt: \\
                  \texttt{\{"alpha\_rate": 0.1, "beta\_rate": 1.0\}} \\
                  (Returned as a Markdown code block)
            \item Disease-informed prompt: \\
                  \texttt{\{"alpha\_rate": 0.1, "beta\_rate": 2.0\}} \\
                  (Returned as a Markdown code block)
        \end{itemize}
\end{itemize}

\section{Computational Requirements}
The experimental pipeline involves extensive MCMC sampling for hierarchical Bayesian models, which represents the primary computational bottleneck. Each model fitting requires 4 parallel chains with 2000 iterations (1000 burn-in + 1000 sampling), performed across multiple cross-validation folds and experimental conditions. LLM API calls add minimal computational overhead compared to the MCMC sampling requirements.

\section{Full Disease-Informed Prompt}
\label{supp:full-disease-prompt}

\begin{listing}[htbp]
\caption{Full prompt for LLM-based Disease-Informed Prior Elicitation.}
\label{lst:supp-disease-prompt}
\begin{lstlisting}
You are a biostatistics expert specializing in oncology clinical trials and Bayesian analysis.

TASK: Provide ONLY rate parameters for exponential priors based on HRPC control arm data.

Clinical Context:
- Disease: Metastatic hormone-resistant prostate cancer (HRPC)
- Treatment: Control arm (placebo/standard care)
- Population: Adult oncology patients
- Study: Multi-center RCT

Model: 
- Each patient i in site j has AE count: y_ij ~ Poisson(lambda_j)
- Site-specific rates: lambda_j ~ Gamma(alpha, beta)  
- REQUIRED: alpha ~ Exponential(rate_alpha), beta ~ Exponential(rate_beta)

IMPORTANT:
- Use your expert knowledge and draw on published clinical trials, empirical data, or established domain knowledge specific to HRPC control arms to set informative (not weakly-informative or non-informative) prior rates.
- Avoid using vague or default values. Base your answer on realistic HRPC control arm data or strong prior experience relevant to typical AE rates in multi-center oncology trials.

RESPOND WITH EXACTLY THIS JSON FORMAT (no markdown, no backticks, no other text):
{
"alpha_rate": number,
"beta_rate": number
}

Note: Exponential(rate) has mean = 1/rate. Rate must be > 0.
\end{lstlisting}
\end{listing}